\documentclass[aps,12pt,
		prl,
		reprint,
		twocolumn,
		superscriptaddress,
		shortbibliography,
		nofootinbib,
		floatfix,
		notitlepage
		]{revtex4-1}

\usepackage{amsmath,amssymb,amsfonts}

\usepackage{natbib}
\usepackage[T1]{fontenc}
\usepackage[latin1]{inputenc}

\usepackage[dvipsnames]{xcolor}

\usepackage{hyperref}
\hypersetup{colorlinks=true,citecolor=Cyan,urlcolor=Red}

\usepackage{mathrsfs}
\usepackage{bbm}
\usepackage{slashed}
\DeclareSymbolFont{rsfs}{U}{rsfs}{m}{n}
\DeclareSymbolFontAlphabet{\mathrsfs}{rsfs}
\usepackage[mathscr]{eucal}	
\usepackage[normalem]{ulem}
\usepackage{verbatim}
\usepackage{bm}

\usepackage{graphicx}

\definecolor{bc}{rgb}{0, 0.7, 0.0}

\newcommand{\be}{\begin{equation}}
\newcommand{\ee}{\end{equation}}
\newcommand{\bi}{\begin{itemize}}
\newcommand{\ei}{\end{itemize}}
\newcommand{\bea}{\begin{eqnarray}}
\newcommand{\eea}{\end{eqnarray}}
\newcommand{\ud}{\mathrm{d}}		

\newcommand{\LCperp}{{\scriptscriptstyle \perp}}

\begin{document}
%


\title{Coherent quantum enhancement of pair production in the null domain}

\author{Anton Ilderton}
\email{anton.ilderton@plymouth.ac.uk}
\affiliation{Centre for Mathematical Sciences, University of Plymouth, Plymouth, PL4 8AA, UK}

\begin{abstract}	
We present an exactly solvable example of coherent quantum interference effects in the creation of electron-positron pairs from the collision of a photon with ultra-short laser pulses.  Being characterised entirely by null, or lightlike, directions, this setup realises an all-optical double-slit in the ``null'' domain, and exhibits features both in common with, and distinct from, a time domain double-slit (Ramsey interferometer). We show that by tailoring the order and amplitude of the pulses one can control signatures of both quantum and classical physics in the produced positron spectrum.
\end{abstract}
\maketitle

The properties of the quantum vacuum allow for all-optical analogues of the double-slit experiment, in which patches of space are polarised by strong fields, effectively creating a diffraction grating through which probe light can be passed to exhibit quantum interference~\cite{King:2013am,King:2013zz}. As well as such `spatial' realisations, it is also possible to realise a temporal, `time domain' double-slit, or Ramsey interferometer~\cite{Ramsey}, based on the quantum vacuum; applying sequences of time-delayed electric fields~\cite{Hebenstreit:2009km,Akkermans:2011yn}, pairs are created from the vacuum via the Schwinger effect~\cite{schwinger51} and their spectra exhibit coherent quantum interference.

Here we combine spatial and temporal realisations of all-optical multiple slits, through pair production stimulated by both strong fields and photonic probes. As both the photon and field will be characterised by null, or lightlike, directions, we refer to the resulting interference as being in the `null' domain. Existing investigations of the Schwinger effect and associated interference have used a variety of versatile techniques, including e.g.~numerical solution of quantum kinetic equations~\cite{Hebenstreit:2009km} and semiclassical approximations~\cite{Akkermans:2011yn,Dumlu:2011rr,Schneider:2018huk}. {Here we will give, to leading order in the fine structure constant, results which are exact in the strong field and of closed form.} We can do so by considering the limit of ultra-short field duration, modelled though delta-function pulses. (See~\cite{Fedotov:2013uja} for delta pulses in Schwinger pair creation.) This will allow us to clearly identify how both quantum effects, e.g.~path-interference, and classical effects, e.g.~post-creation acceleration, appear in observables, giving insight into the impact which the spacetime distribution of multiple pulses has on the produced pair spectrum~\cite{QLV}.

Our calculations have the advantage of supplying {explicit} and easily interpretable results which exhibit interference effects common in many tunnelling phenomena, for example strong-field ionisation~\cite{Lindner,control,ionisation}.  {With an eye to experiment, we remark that analogues of multiple slit interference persist in stimulated pair production even when pulses are not ultra-short~\cite{Heinzl:2010vg}, and also that short, intense pulses of femto- and atto-second duration be constructed by a variety of methods~\cite{RES,XUV}.} Further, stimulated pair production is more immediately realisable, experimentally~\cite{SLAC-E144}, than the Schwinger effect (which requires extreme field strengths even with optimisation of the field profile~\cite{Schutzhold:2008pz,Dunne:2009gi,DiPiazza:2009py,Bulanov:2010ei,Kohlfurst:2012rb,Gonoskov:2013ada,Hebenstreit:2014lra}) and indeed will be investigated in upcoming experiments~\cite{Abramowicz:2019gvx,E320}.

\begin{figure}[t!]
\raisebox{30pt}{\includegraphics[width=0.3\columnwidth]{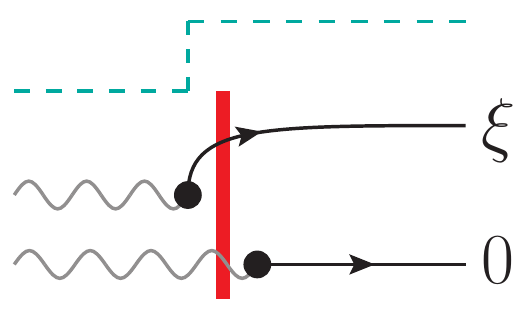}}\, \, \includegraphics[width=0.6\columnwidth]{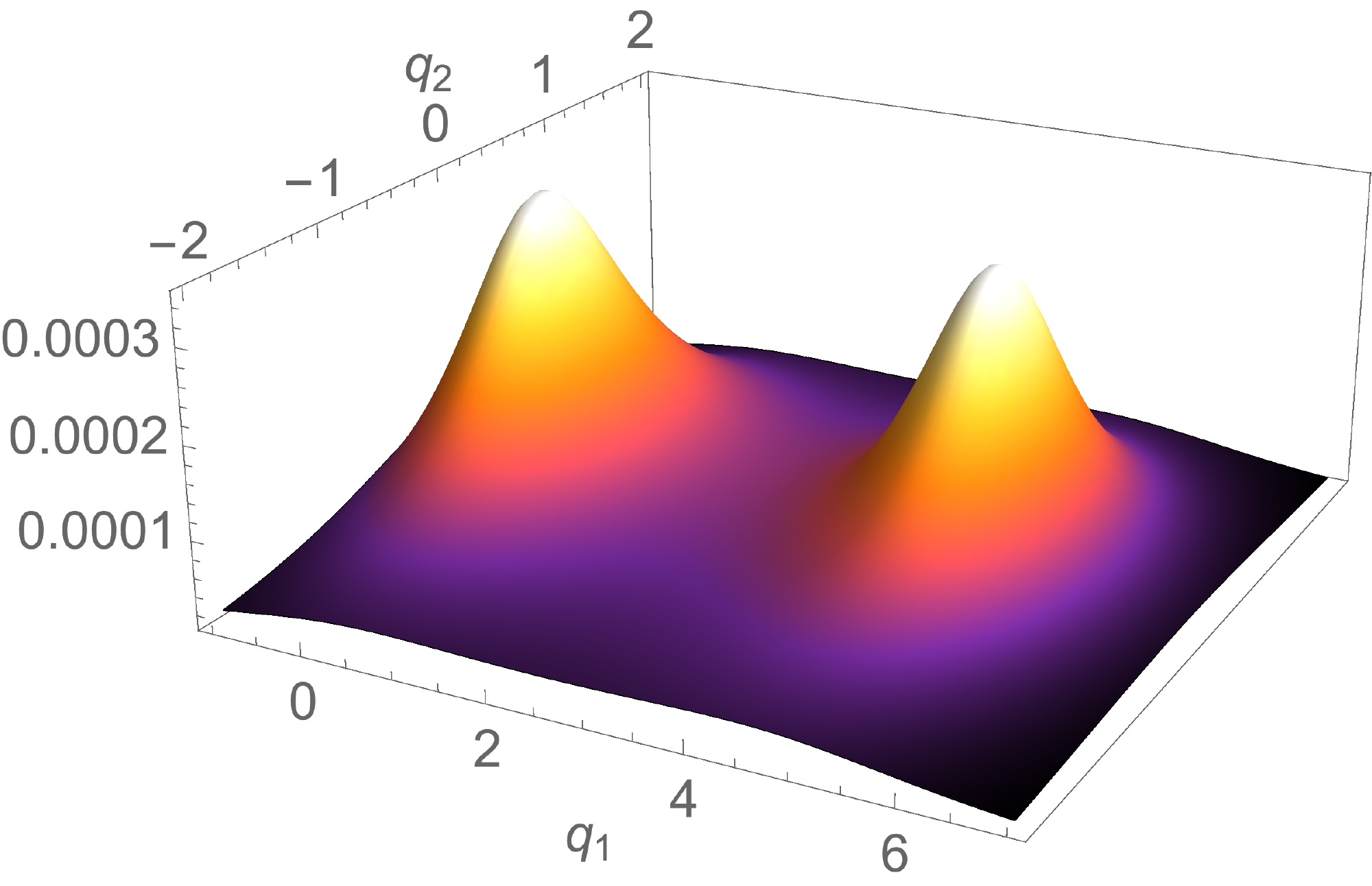}
\caption{\label{FIG:1}
\textit{Left}: in an ultra-short pulse (red line), positrons created in the rise of the field are accelerated, yielding a spectral peak at momentum $a_\LCperp=\{m \xi, 0\}$, while those created in the fall of the field are not, leading to a peak at zero momentum. Dashed lines illustrate the potential. {\textit{Right}: the spectrum for $l_\LCperp=0$, $u=1/2$, $\xi=5$ as a function of transverse positron momenta $\{q_1,q_2\}$ in units of~$m$ ($q_1$ is the momentum in the field polarisation direction)}, showing the two spectral peaks.}
\end{figure}

Our starting point is the {quantum mechanical probability} of electron-positron pair creation from a probe photon, momentum $l_\mu$,  in the presence of classical electric and magnetic fields modelling a {strong} laser pulse. This will be a plane wave travelling in the $-z$ direction, thus depending on `lightfront time' $\phi:= t +z$. The transverse electromagnetic fields of the wave are described by the potential $eA_\mu \equiv a_\mu = (0,a_1(\phi),a_2(\phi),0)$ with $eE^1 = eB^2 = -a_1'$, $eE^2= - eB^1 = -a_2'$. We consider a linearly polarised field, $a_2=0$, of ultra-short duration such that the electric field becomes a delta function, so $a_1 \to m \xi H(\phi)$, a Heaviside step function of (dimensionless) height~$\xi$. This parameterisation corresponds to taking the simultaneous limit of high field strength and short pulse duration~\cite{Jag}, such that the total work done on a particle traversing the field remains fixed -- this is $m\xi$~\cite{Dinu:2012tj}. (As such $\xi$ matches the usual definition of the intensity parameter in laser-matter interactions~\cite{RitusReview,DiPiazza:2011tq}.)

\begin{figure*}[t!]
{\includegraphics[width=0.4\textwidth]{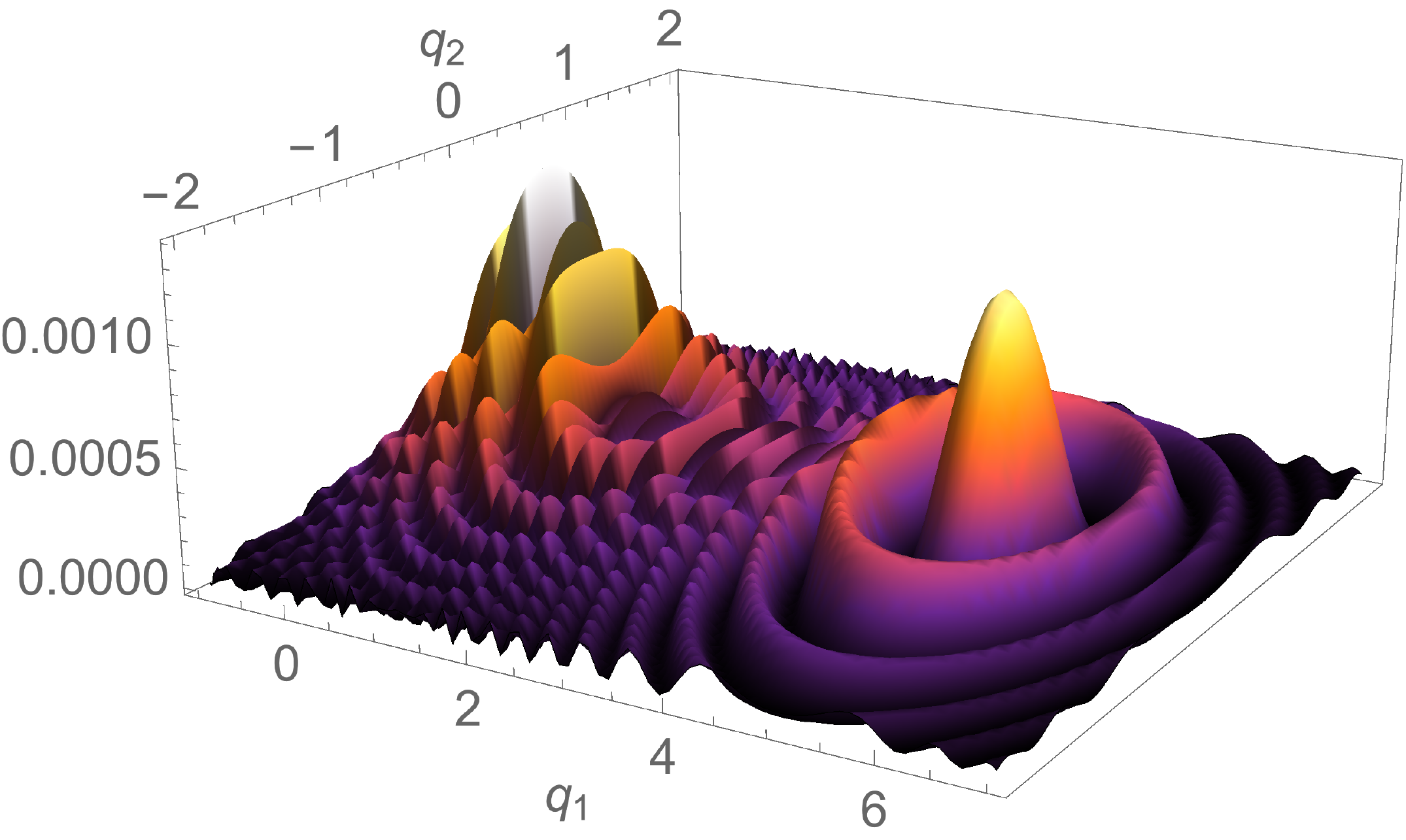}}\qquad
\raisebox{20pt}{\includegraphics[width=0.29\textwidth]{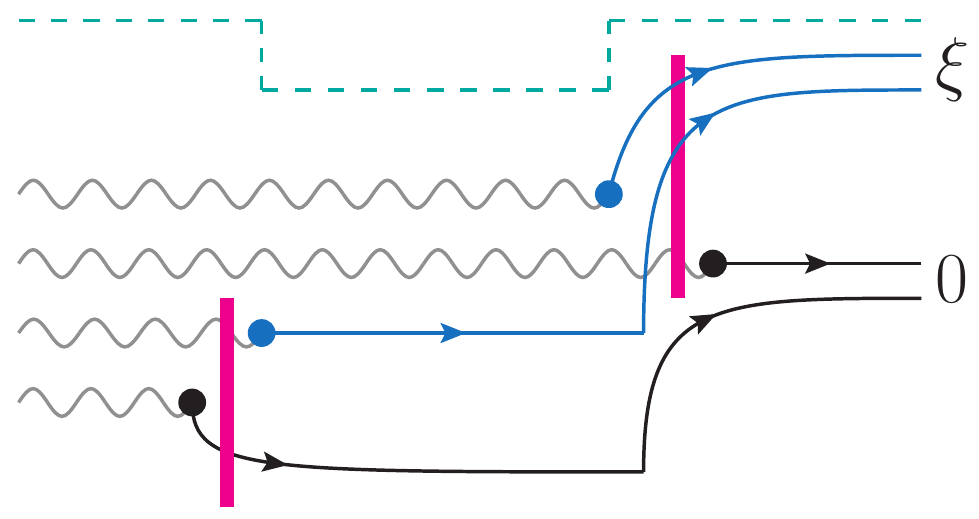}}
\caption{\label{FIG:2} {Coherent quantum interference in the spectrum of positrons produced in a sequence of two delta-pulses of opposite sign, with other parameters as for Fig.~\ref{FIG:1}. The negative amplitude pulse occurs causally before that with positive amplitude}, {and $\theta=1$ here.} The two spectral peaks from the single pulse case in Fig.~\ref{FIG:1} persist (left), amplified by coherent interference arising via two possible creation + acceleration events, illustrated on the right.}
\end{figure*}

The calculation of the pair creation probability employs the well-known Volkov solutions~\cite{volkov35} and is given in~\cite{Jag}. We also use the LSZ reduction of~\cite{Kibble:1965zza} which accounts for the asymptotically nonvanishing potential~$a_1$ and has the advantage of allowing us to parameterise the  probability with the produced positron's \textit{physical}, final momentum $q_\LCperp \equiv \{q_1,q_2\}$, transverse to the laser propagation direction, and its longitudinal `lightfront' momentum $u:= n\cdot q/n \cdot l$, where $n_\mu=(1,0,0,1)$. From this we define a second on-shell momentum $\bar{\pi}^\mu := q^\mu - a^\mu + n^\mu (2q\cdot a-a^2)/2n\cdot q$ which is, using the exact solution of the Lorentz force equation, equal to the classical \textit{initial} momentum of the positron, had it traversed the delta function and been accelerated to final momentum $q_\mu$~\cite{Dinu:2012tj}. In terms of these momenta the pair production probability {$\mathbb P$ is, for $\alpha$ the fine-structure constant},
\be\label{DC-PAR-1}
	\mathbb{P} = \frac{\alpha m^2}{4\pi^2} \int\! \ud^2 q_\LCperp\! \int\limits_0^1\!\frac{\ud u}{u}  \, (1-u) \, F[q,\bar\pi,\xi]\;,
\ee
where the key quantity $F$ is, for $h := 1/2 - 1/4 u(1-u)$, 
\be
\label{DC-PAR-0}
	F[q, \bar\pi, \xi] := \frac{1}{(l\!\cdot\!  q)^2} + \frac{1}{(l\!\cdot\! \bar\pi)^2}  - \frac{2}{l\!\cdot\! q\, l\!\cdot\! {\bar \pi}}\big(1+\xi^2 h\big) \;.
\ee
This expression is similar to that obtained for scattering of a particle off an instantaneous kick~\cite{Peskin:1995ev}, but with additional field-dependent structure.

The produced positron momentum spectrum is given by stripping the integrals from (\ref{DC-PAR-1}). We show the spectrum in Fig.~\ref{FIG:1} for $l_\LCperp=0$, describing a head-on collision between the photon and field, and $u=1/2$, the symmetric point corresponding to the produced positron and electron each carrying half of the initial photon's lightfront momentum. The spectrum exhibits \textit{two} peaks at momenta $q_\LCperp = 0$ and $q_\LCperp = a_\LCperp$. These arise primarily from the first two terms in $F$, see~(\ref{DC-PAR-0}), while the final `cross' term contributes more broadly across the whole spectrum. To explain the double-peak structure, we first note that if the sign of the delta-pulse is swapped, then the second peak switches to $q_\LCperp=-a_\LCperp$. Assume then that the dominant contribution to the spectrum comes from particles \textit{created} at zero transverse momentum. (This is natural since the sum of the pair's transverse momenta is conserved~\cite{RitusReview,DiPiazza:2011tq,Seipt:2017ckc} and equal to $l_\LCperp$, which is zero here.) Then positrons created in the rapid rise of an ultra-strong, ultra-short pulse see and are accelerated by essentially the whole field, picking up the full possible transverse momentum $a_\LCperp$ from it. Hence the spectral peak at $q_\LCperp = a_\LCperp$. Positrons created in the fall of the field, on the other hand, see little of it and so acquire little momentum after creation. Hence the spectral peak at $q_\LCperp = 0$. This interpretation is consistent with the change in the spectrum when the pulse changes sign, for then the former source of positrons picks up $-a_\LCperp$ when accelerated by the field.

We remark that time-domain spectra for Schwinger pair production in time-dependent electric fields typically exhibit only a \textit{single} spectral peak at zero momentum~\cite{Hebenstreit:2009km,Akkermans:2011yn,Dumlu:2011rr}. Since the semiclassical calculations used there require the electric field strength to be small (much lower than the Schwinger field $E_S=m^2/e$), we compare by taking $\xi\ll 1$. In this case our two spectral peaks coalesce into a single structure at $q_\LCperp=0$, similar to the time domain results. Thus we can make contact with the literature by calculating perturbatively in $\xi$, which will be useful below, but our calculations already reveal structures beyond existing results.

With this, we can turn to the investigation of quantum interference effects.  Consider two delta pulses {of opposite sign and} separated by a distance $2\Delta\phi$ in lightfront time~$\phi$. We will see that there is a strong causal aspect to our results, so that we take the `positive' delta-pulse (as above) to lie at the larger value of $\phi$. We can again calculate the spectrum exactly~\cite{Jag}; it is given by making the replacement, in the single pulse result (\ref{DC-PAR-1}),
\be\label{P-PAR-UTAN-DC}
	F[q, \bar \pi, \xi] \longrightarrow 4\sin^2(\Theta)\, F[q, \bar \pi, \xi] \;,
\ee
where $\Theta = \Delta \phi\,\, l \cdot  \!\bar\pi /n\!\cdot  l\,(1-u)$ arises in the calculation as an accumulated phase~\cite{Akkermans:2011yn} between the two pulses. Thus the effect of adding a second pulse of opposite sign is, without approximation, to \textit{coherently} enhance the positron spectrum through a double slit interference pattern. This is shown in Fig.~\ref{FIG:2} (left panel), for the same~$\xi$ as in Fig.~\ref{FIG:1}; the interference fringes are clearly visible {and are controlled by the combined separation/energy parameter $\theta = \Delta\phi\, m^2 /n\!\cdot\! l$ which appears in the interference angle $\Theta$.} Now {we observe that $\Theta$ may be written
\be
	\Theta = \frac{u}{2(1-u)} l_\mu \int_{-\Delta\phi}^{\Delta\phi}\!\ud\varphi\, \frac{\bar\pi^\mu}{n \!\cdot\!  q} \equiv \frac{u}{2(1-u)} \, l \!\cdot\! \delta x \;.
\ee
The integral $\delta x^\mu$ is, using the Lorentz force equation, exactly equal to the \textit{change in position} of a classical positron, created in the first pulse, during the elapsed lightfront time between the two pulses.} This suggests that the physical origin of the interference pattern in~(\ref{P-PAR-UTAN-DC}) is path-interference between trajectories of produced positrons. Using the above physical picture, we can easily identify the dominant sources of interference. As illustrated in Fig~\ref{FIG:2}, positrons created in the rise of the first peak are accelerated by the whole field, picking up $-a_\LCperp$ in transverse momentum at the first delta, but then $+a_\LCperp$ at the second, ending with zero transverse momentum. Positrons created in the fall of the second peak receive no momentum, as before. These \textit{two} `creation and acceleration' events both source a spectral peak at $q_\LCperp=0$, and interfere. Positrons created in the fall/rise of the first/second delta, on the other hand, pick up only $+a_\LCperp$ in transverse momentum. These are the two events which interfere and source the second spectral peak at $q_\LCperp= a_\LCperp$. Further, if we exchange the signs of the deltas, we find that the spectral peaks appear at $q_\LCperp=0$ and, now, $q_\LCperp=-a_\LCperp$; this is consistent with our interpretation, because changing the signs of the pulses has the same effect as exchanging their time \textit{ordering}, and massive particles always propagate from smaller to larger $\phi$~\cite{Dirac:1955uv,Brodsky:1997de,Heinzl:2000ht}. Hence  the positrons created in the fall/rise of the first/second peak now pick up $-a_\LCperp$ when accelerated by the field.

The interplay of quantum interference and classical dynamics in our results has an interesting consequence when we consider a sequence of two pulses of the \text{same sign}. From studies of Schwinger pair production which realise a double-slit in the \textit{time} domain, coherent enhancement would \text{not} be expected in this case~\cite{Akkermans:2011yn}.  {However, those results hold in the semiclassical approximation, which we can here go beyond to identify new effects.} 

\begin{figure}[t!!]
{\includegraphics[width=0.4\textwidth]{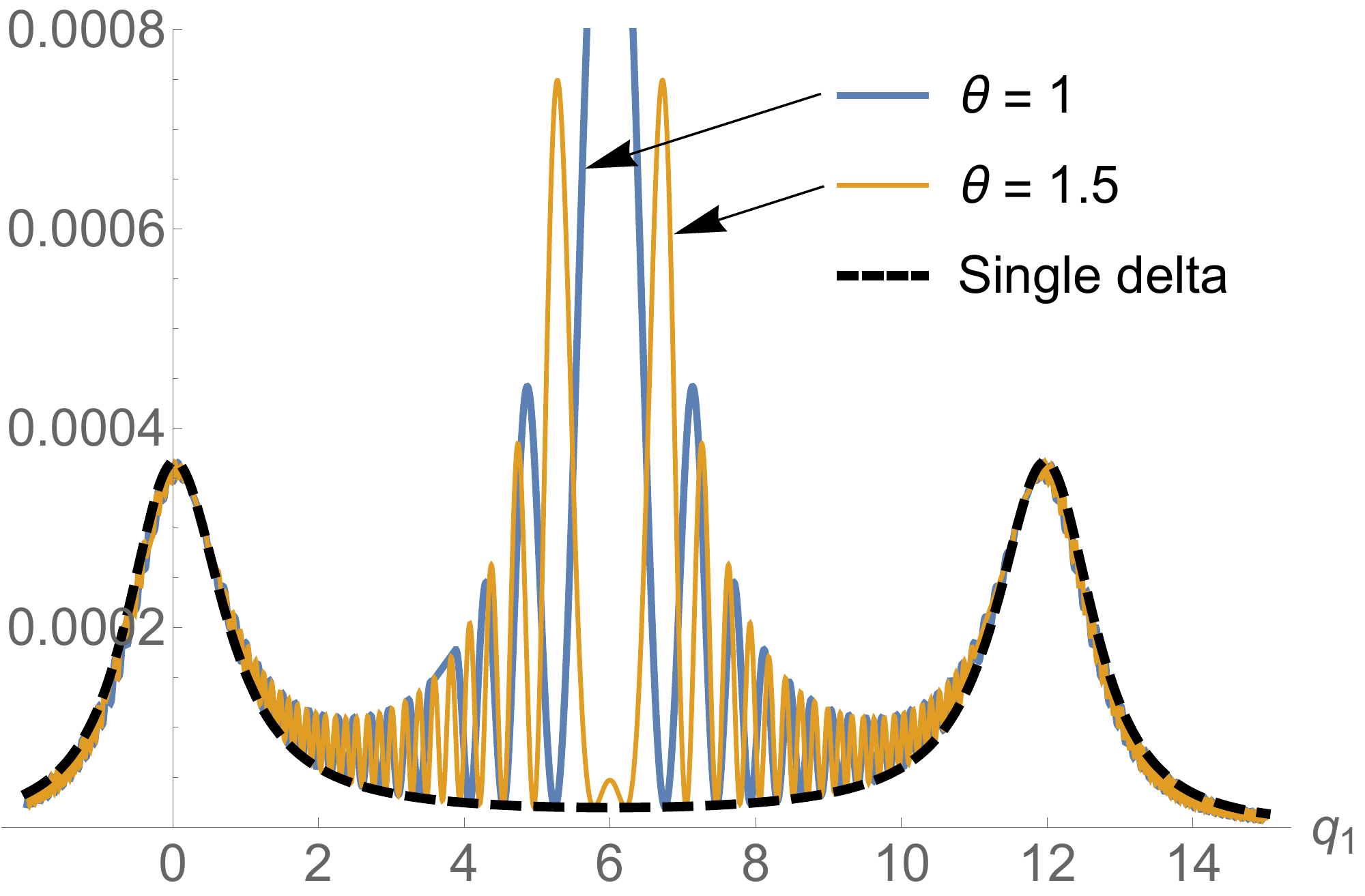}} \\
\hspace{-20pt}{\includegraphics[width=0.39\textwidth]{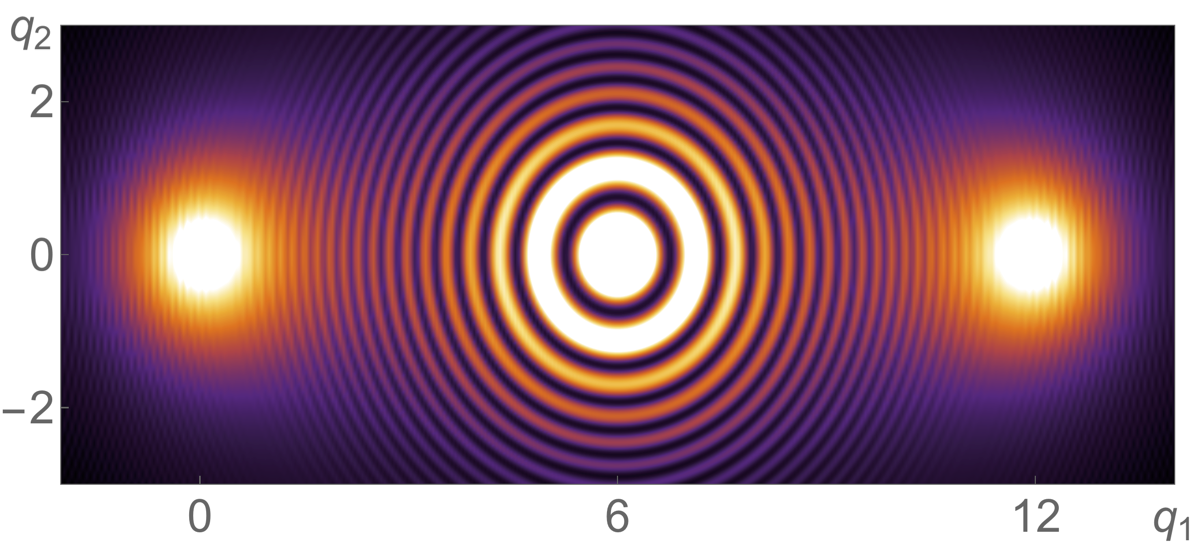}}
{\includegraphics[width=0.3\textwidth]{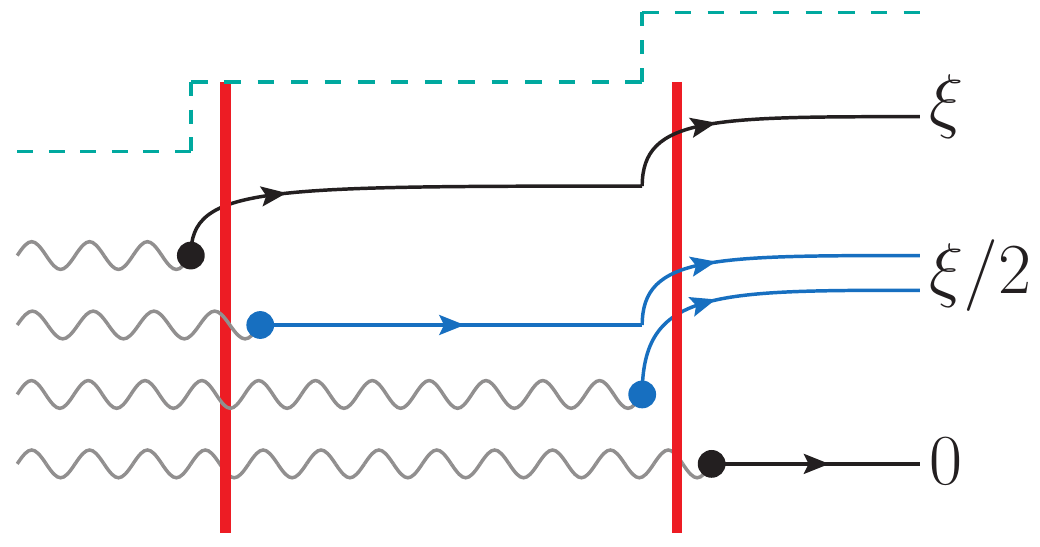}}
\caption{\label{FIG:3} In a sequence of two pulses with the same sign, the spectrum exhibits three peaks (upper panel, $\xi=12$, $\theta\in\{1,1.5\}$ and $q_2=0$).  The outer peaks align with those of the single pulse, and receive only mild interference effects. The third peak lies halfway between these two and is coherently enhanced by a $4\sin^2$ factor. The middle panel shows {the same} spectrum as a function of $q_1$ and $q_2$ ($\theta=1$); the outer peaks are smooth, while interference fringes clearly dominate the middle peak. The dominant pair creation events sourcing the peaks and interference are illustrated in the bottom panel.}
\end{figure}

We take two delta pulses of the same sign and (for reasons which will become clear) each doing \textit{half} the work of the previous case, so $\xi\to \xi/2$.  There is now only one dominant contribution to the zero momentum part of the spectrum, namely pairs created after the second pulse, see  Fig.~\ref{FIG:3}. Similarly, only pairs created before the first pulse, and then accelerated twice, source a spectral peak at $q_\LCperp = a_\LCperp$. Hence we expect peaks at $q_\LCperp=0$ and $q_\LCperp=a_\LCperp$ as before, without interference or enchancement. This is as expected from the time-domain results. However, positrons created in the fall of the first pulse, or the rise of the second, are both accelerated to $a_\LCperp/2$; hence, if our physical interpretation is correct, we should expect a \textit{third} spectral peak, between the others, with coherent enhancement. To verify this, define ${\hat \pi}_\mu$ to be equal to ${\bar \pi}_\mu$ but with $\xi \to \xi/2$; the positron spectrum and production probability in this case are now given by \textit{adding to} $F[q,\bar\pi,\xi]$ in (\ref{DC-PAR-1}) the term
\begin{align}\label{FFF}
	2\sin^2(\Theta) \Big[ F[q, \hat\pi, \tfrac12 \xi] + F[\bar\pi, \hat \pi, \tfrac12 \xi]-F[q,\bar\pi,\xi]\Big] \;.
\end{align}
Writing out the term multiplying $2\sin^2(\Theta)$ explicitly, we find the following `additive' correction to the single pulse spectrum: there is a quadratic term $2/l \cdot \hat\pi^2$, yielding a peak at $q_\LCperp = a_\LCperp/2$, multiplied by, overall, the coherent interference factor $4\sin^2(\Theta)$. The remaining terms are all `cross', going like $1/(l \cdot \hat\pi l \cdot q)$ etc, which contribute to a lesser extent across all peaks. Thus, in agreement with our predictions and as confirmed in Fig.~\ref{FIG:3}, the single-pulse peaks persist, showing only weak interference effects, while the new peak is coherently enhanced relative to that in a single pulse of intensity parameter $\xi/2$~\cite{note1}.

It follows that by changing the order and signs of several delta pulses we can exert a degree of control on the spectrum (beyond simply adjusting the separation $\Delta\phi$ to change the frequency of oscillations in $\Theta$): we can arrange for interference to appear in one or multiple peaks, and not others. For example, if we have two ordered pulses characterised by $-\xi/2$ and $+\xi$, then the spectrum will exhibit single-pulse peaks at $q_\LCperp=0$ and $a_\LCperp/2$, with little interference, and a third peak with coherent enhancement at $q_\LCperp = a_\LCperp$. Adding further pulses can increase the degree of enhancement and introduce additional parameters which affect the spectrum. To illustrate this we consider a final example of four equally spaced pulses of alternating signs ($-$+$-$+). In the time domain, this setup yields a 4-slit interference pattern overlaying the single peak spectrum, but here the structure is richer. We find that the single pulse spectrum is multiplied by
\be\label{Theta01}
	4^2\sin^2\Theta \cos^2 (\Theta+\Theta_0) \;,
\ee
in which the new parameter $\Theta_0$, equal to $\Theta$ evaluated at zero field, arises from free particle propagation between the second and third deltas, whereas $ \Theta$ arises as above from propagation between e.g.~the first and second. To recover the time domain results we can again take the weak field limit $\xi\ll1$. In this limit $\Theta \to \Theta_0$ and so our 4-pulse factor (\ref{Theta01}) reduces to
\be
	4^2\sin^2(\Theta_0) \cos^2 (2\,\Theta_0) = \frac{\sin^2 (4\,\Theta_0)}{\cos^2(\Theta_0)} \;,
\ee
which is precisely the expected 4-slit Fabry-Perot interference pattern. This shows nicely that multiple-slit interference underlines pair production in weak fields, but that {in the strong field, $\xi>1$, regime, which is experimentally accessible~\cite{SLAC-E144,Abramowicz:2019gvx,E320}}, there is even more structure. This structure is fully explorable in our model.

To conclude, we have described quantum interference effects in an all-optical double-slit setup, in which pairs are produced from the quantum vacuum through the interaction of a photon with ultra-short laser pulses. By idealising the laser fields as delta-function pulses we have been able to give exact, closed-form results for the produced positron spectrum which explicitly demonstrate coherent quantum enhancement when multiple pulses are present. The produced positron spectra have a depth of structure not noted before, and which goes beyond existing semiclassical descriptions, yet admits a simple physical interpretation in terms of two-path interference~\cite{Glory}, and the classical physics (charge acceleration) along those paths. {It would be interesting to obtain the analogous results for the Schwinger effect in the time-domain, though this remains challenging analytically. (See e.g.~\cite{Dong:2017vse} for numerical results on two \textit{spatially} separated pulses of the same sign.)} We hope also that our results will be useful in other areas -- it has been shown for example that understanding path interference can resolve discrepancies between the theory and experiment~\cite{Huismans2,Glory} of tunnelling phenomena in photoelectron holography~\cite{Huismans1,Porat}.

We have also shown that the location and degree of interference effects in positron spectra can be controlled~\cite{Dumlu:2011rr,QLV} by changing the number and amplitude of the pulses. Armed with this understanding, {it will be interesting to establish in future work the extent to which spectral control can be achieved at upcoming experiments~\cite{Abramowicz:2019gvx,E320}. This requires establishing the role of e.g.~transverse size effects on pairs, post creation~\cite{Harvey:2016wte}; these can be mitigated by using short pulses~\cite{Harvey:2016uiy} (in which interference does persist~\cite{Heinzl:2010vg}), which would be the goal.} We also note that interference has been used, experimentally, in laser-based analogue computers~\cite{Bigourd} with pulses of femtosecond duration, yet agreement with theory based on delta-function fields was nevertheless excellent. This also hints at the possibility of using pair production interference as an analogue computer.

\textit{A.I.~thanks Gerald Dunne for many useful discussions. A.I.~is supported by the EPSRC, grant EP/S010319/1.}


\begin{thebibliography}{99}%
\makeatletter
\providecommand \@ifxundefined [1]{%
 \@ifx{#1\undefined}
}%
\providecommand \@ifnum [1]{%
 \ifnum #1\expandafter \@firstoftwo
 \else \expandafter \@secondoftwo
 \fi
}%
\providecommand \@ifx [1]{%
 \ifx #1\expandafter \@firstoftwo
 \else \expandafter \@secondoftwo
 \fi
}%
\providecommand \natexlab [1]{#1}%
\providecommand \enquote  [1]{``#1''}%
\providecommand \bibnamefont  [1]{#1}%
\providecommand \bibfnamefont [1]{#1}%
\providecommand \citenamefont [1]{#1}%
\providecommand \href@noop [0]{\@secondoftwo}%
\providecommand \href [0]{\begingroup \@sanitize@url \@href}%
\providecommand \@href[1]{\@@startlink{#1}\@@href}%
\providecommand \@@href[1]{\endgroup#1\@@endlink}%
\providecommand \@sanitize@url [0]{\catcode `\\12\catcode `\$12\catcode
  `\&12\catcode `\#12\catcode `\^12\catcode `\_12\catcode `\%12\relax}%
\providecommand \@@startlink[1]{}%
\providecommand \@@endlink[0]{}%
\providecommand \url  [0]{\begingroup\@sanitize@url \@url }%
\providecommand \@url [1]{\endgroup\@href {#1}{\urlprefix }}%
\providecommand \urlprefix  [0]{URL }%
\providecommand \Eprint [0]{\href }%
\providecommand \doibase [0]{http://dx.doi.org/}%
\providecommand \selectlanguage [0]{\@gobble}%
\providecommand \bibinfo  [0]{\@secondoftwo}%
\providecommand \bibfield  [0]{\@secondoftwo}%
\providecommand \translation [1]{[#1]}%
\providecommand \BibitemOpen [0]{}%
\providecommand \bibitemStop [0]{}%
\providecommand \bibitemNoStop [0]{.\EOS\space}%
\providecommand \EOS [0]{\spacefactor3000\relax}%
\providecommand \BibitemShut  [1]{\csname bibitem#1\endcsname}%
\let\auto@bib@innerbib\@empty
\bibitem [{\citenamefont {King}\ \emph
  {et~al.}(2010{\natexlab{a}})\citenamefont {King}, \citenamefont {Di~Piazza},\
  and\ \citenamefont {Keitel}}]{King:2013am}%
  \BibitemOpen
  \bibfield  {author} {\bibinfo {author} {\bibfnamefont {B.}~\bibnamefont
  {King}}, \bibinfo {author} {\bibfnamefont {A.}~\bibnamefont {Di~Piazza}}, \
  and\ \bibinfo {author} {\bibfnamefont {C.~H.}\ \bibnamefont {Keitel}},\
  }\href {\doibase 10.1038/nphoton.2009.261} {\bibfield  {journal} {\bibinfo
  {journal} {Nature Photon.}\ }\textbf {\bibinfo {volume} {4}},\ \bibinfo
  {pages} {92} (\bibinfo {year} {2010}{\natexlab{a}})},\ \Eprint
  {http://arxiv.org/abs/1301.7038} {arXiv:1301.7038 [physics.optics]}
  \BibitemShut {NoStop}%
\bibitem [{\citenamefont {King}\ \emph
  {et~al.}(2010{\natexlab{b}})\citenamefont {King}, \citenamefont {Di~Piazza},\
  and\ \citenamefont {Keitel}}]{King:2013zz}%
  \BibitemOpen
  \bibfield  {author} {\bibinfo {author} {\bibfnamefont {B.}~\bibnamefont
  {King}}, \bibinfo {author} {\bibfnamefont {A.}~\bibnamefont {Di~Piazza}}, \
  and\ \bibinfo {author} {\bibfnamefont {C.~H.}\ \bibnamefont {Keitel}},\
  }\href {\doibase 10.1103/PhysRevA.82.032114} {\bibfield  {journal} {\bibinfo
  {journal} {Phys. Rev.}\ }\textbf {\bibinfo {volume} {A82}},\ \bibinfo {pages}
  {032114} (\bibinfo {year} {2010}{\natexlab{b}})},\ \Eprint
  {http://arxiv.org/abs/1301.7008} {arXiv:1301.7008 [physics.optics]}
  \BibitemShut {NoStop}%
\bibitem [{\citenamefont {Ramsey}(1950)}]{Ramsey}%
  \BibitemOpen
  \bibfield  {author} {\bibinfo {author} {\bibfnamefont {N.~F.}\ \bibnamefont
  {Ramsey}},\ }\href {\doibase 10.1103/PhysRev.78.695} {\bibfield  {journal}
  {\bibinfo  {journal} {Phys. Rev.}\ }\textbf {\bibinfo {volume} {78}},\
  \bibinfo {pages} {695} (\bibinfo {year} {1950})}\BibitemShut {NoStop}%
\bibitem [{\citenamefont {Hebenstreit}\ \emph {et~al.}(2009)\citenamefont
  {Hebenstreit}, \citenamefont {Alkofer}, \citenamefont {Dunne},\ and\
  \citenamefont {Gies}}]{Hebenstreit:2009km}%
  \BibitemOpen
  \bibfield  {author} {\bibinfo {author} {\bibfnamefont {F.}~\bibnamefont
  {Hebenstreit}}, \bibinfo {author} {\bibfnamefont {R.}~\bibnamefont
  {Alkofer}}, \bibinfo {author} {\bibfnamefont {G.~V.}\ \bibnamefont {Dunne}},
  \ and\ \bibinfo {author} {\bibfnamefont {H.}~\bibnamefont {Gies}},\ }\href
  {\doibase 10.1103/PhysRevLett.102.150404} {\bibfield  {journal} {\bibinfo
  {journal} {Phys. Rev. Lett.}\ }\textbf {\bibinfo {volume} {102}},\ \bibinfo
  {pages} {150404} (\bibinfo {year} {2009})},\ \Eprint
  {http://arxiv.org/abs/0901.2631} {arXiv:0901.2631 [hep-ph]} \BibitemShut
  {NoStop}%
\bibitem [{\citenamefont {Akkermans}\ and\ \citenamefont
  {Dunne}(2012)}]{Akkermans:2011yn}%
  \BibitemOpen
  \bibfield  {author} {\bibinfo {author} {\bibfnamefont {E.}~\bibnamefont
  {Akkermans}}\ and\ \bibinfo {author} {\bibfnamefont {G.~V.}\ \bibnamefont
  {Dunne}},\ }\href {\doibase 10.1103/PhysRevLett.108.030401} {\bibfield
  {journal} {\bibinfo  {journal} {Phys. Rev. Lett.}\ }\textbf {\bibinfo
  {volume} {108}},\ \bibinfo {pages} {030401} (\bibinfo {year} {2012})},\
  \Eprint {http://arxiv.org/abs/1109.3489} {arXiv:1109.3489 [hep-th]}
  \BibitemShut {NoStop}%
\bibitem [{\citenamefont {Schwinger}(1951)}]{schwinger51}%
  \BibitemOpen
  \bibfield  {author} {\bibinfo {author} {\bibfnamefont {J.}~\bibnamefont
  {Schwinger}},\ }\href@noop {} {\bibfield  {journal} {\bibinfo  {journal}
  {Phys. Rev.}\ }\textbf {\bibinfo {volume} {82}},\ \bibinfo {pages} {664}
  (\bibinfo {year} {1951})}\BibitemShut {NoStop}%
\bibitem [{\citenamefont {Dumlu}\ and\ \citenamefont
  {Dunne}(2011)}]{Dumlu:2011rr}%
  \BibitemOpen
  \bibfield  {author} {\bibinfo {author} {\bibfnamefont {C.~K.}\ \bibnamefont
  {Dumlu}}\ and\ \bibinfo {author} {\bibfnamefont {G.~V.}\ \bibnamefont
  {Dunne}},\ }\href {\doibase 10.1103/PhysRevD.83.065028} {\bibfield  {journal}
  {\bibinfo  {journal} {Phys. Rev.}\ }\textbf {\bibinfo {volume} {D83}},\
  \bibinfo {pages} {065028} (\bibinfo {year} {2011})},\ \Eprint
  {http://arxiv.org/abs/1102.2899} {arXiv:1102.2899 [hep-th]} \BibitemShut
  {NoStop}%
\bibitem [{\citenamefont {Schneider}\ \emph {et~al.}(2018)\citenamefont
  {Schneider}, \citenamefont {Torgrimsson},\ and\ \citenamefont
  {Sch{\"u}tzhold}}]{Schneider:2018huk}%
  \BibitemOpen
  \bibfield  {author} {\bibinfo {author} {\bibfnamefont {C.}~\bibnamefont
  {Schneider}}, \bibinfo {author} {\bibfnamefont {G.}~\bibnamefont
  {Torgrimsson}}, \ and\ \bibinfo {author} {\bibfnamefont {R.}~\bibnamefont
  {Sch{\"u}tzhold}},\ }\href {\doibase 10.1103/PhysRevD.98.085009} {\bibfield
  {journal} {\bibinfo  {journal} {Phys. Rev.}\ }\textbf {\bibinfo {volume}
  {D98}},\ \bibinfo {pages} {085009} (\bibinfo {year} {2018})},\ \Eprint
  {http://arxiv.org/abs/1806.00943} {arXiv:1806.00943 [hep-th]} \BibitemShut
  {NoStop}%
\bibitem [{\citenamefont {Fedotov}\ and\ \citenamefont
  {Mironov}(2013)}]{Fedotov:2013uja}%
  \BibitemOpen
  \bibfield  {author} {\bibinfo {author} {\bibfnamefont {A.~M.}\ \bibnamefont
  {Fedotov}}\ and\ \bibinfo {author} {\bibfnamefont {A.~A.}\ \bibnamefont
  {Mironov}},\ }\href {\doibase 10.1103/PhysRevA.88.062110} {\bibfield
  {journal} {\bibinfo  {journal} {Phys. Rev.}\ }\textbf {\bibinfo {volume}
  {A88}},\ \bibinfo {pages} {062110} (\bibinfo {year} {2013})},\ \Eprint
  {http://arxiv.org/abs/1310.7258} {arXiv:1310.7258 [hep-ph]} \BibitemShut
  {NoStop}%
\bibitem [{\citenamefont {Lv}\ \emph {et~al.}(2018)\citenamefont {Lv},
  \citenamefont {Su},\ and\ \citenamefont {Grobe}}]{QLV}%
  \BibitemOpen
  \bibfield  {author} {\bibinfo {author} {\bibfnamefont {Q.~Z.}\ \bibnamefont
  {Lv}}, \bibinfo {author} {\bibfnamefont {Q.}~\bibnamefont {Su}}, \ and\
  \bibinfo {author} {\bibfnamefont {R.}~\bibnamefont {Grobe}},\ }\href
  {\doibase 10.1103/PhysRevLett.121.183606} {\bibfield  {journal} {\bibinfo
  {journal} {Phys. Rev. Lett.}\ }\textbf {\bibinfo {volume} {121}},\ \bibinfo
  {pages} {183606} (\bibinfo {year} {2018})}\BibitemShut {NoStop}%
\bibitem [{\citenamefont {Lindner}\ \emph {et~al.}(2005)\citenamefont
  {Lindner}, \citenamefont {Sch\"atzel}, \citenamefont {Walther}, \citenamefont
  {Baltu\ifmmode~\check{s}\else \v{s}\fi{}ka}, \citenamefont {Goulielmakis},
  \citenamefont {Krausz}, \citenamefont {Milo\ifmmode \check{s}\else
  \v{s}\fi{}evi\ifmmode~\acute{c}\else \'{c}\fi{}}, \citenamefont {Bauer},
  \citenamefont {Becker},\ and\ \citenamefont {Paulus}}]{Lindner}%
  \BibitemOpen
  \bibfield  {author} {\bibinfo {author} {\bibfnamefont {F.}~\bibnamefont
  {Lindner}}, \bibinfo {author} {\bibfnamefont {M.~G.}\ \bibnamefont
  {Sch\"atzel}}, \bibinfo {author} {\bibfnamefont {H.}~\bibnamefont {Walther}},
  \bibinfo {author} {\bibfnamefont {A.}~\bibnamefont
  {Baltu\ifmmode~\check{s}\else \v{s}\fi{}ka}}, \bibinfo {author}
  {\bibfnamefont {E.}~\bibnamefont {Goulielmakis}}, \bibinfo {author}
  {\bibfnamefont {F.}~\bibnamefont {Krausz}}, \bibinfo {author} {\bibfnamefont
  {D.~B.}\ \bibnamefont {Milo\ifmmode \check{s}\else
  \v{s}\fi{}evi\ifmmode~\acute{c}\else \'{c}\fi{}}}, \bibinfo {author}
  {\bibfnamefont {D.}~\bibnamefont {Bauer}}, \bibinfo {author} {\bibfnamefont
  {W.}~\bibnamefont {Becker}}, \ and\ \bibinfo {author} {\bibfnamefont {G.~G.}\
  \bibnamefont {Paulus}},\ }\href {\doibase 10.1103/PhysRevLett.95.040401}
  {\bibfield  {journal} {\bibinfo  {journal} {Phys. Rev. Lett.}\ }\textbf
  {\bibinfo {volume} {95}},\ \bibinfo {pages} {040401} (\bibinfo {year}
  {2005})}\BibitemShut {NoStop}%
\bibitem [{\citenamefont {Maxwell}\ and\ \citenamefont
  {Faria}(2016)}]{control}%
  \BibitemOpen
  \bibfield  {author} {\bibinfo {author} {\bibfnamefont {A.~S.}\ \bibnamefont
  {Maxwell}}\ and\ \bibinfo {author} {\bibfnamefont {C.~F. d.~M.}\ \bibnamefont
  {Faria}},\ }\href {\doibase 10.1103/PhysRevLett.116.143001} {\bibfield
  {journal} {\bibinfo  {journal} {Phys. Rev. Lett.}\ }\textbf {\bibinfo
  {volume} {116}},\ \bibinfo {pages} {143001} (\bibinfo {year}
  {2016})}\BibitemShut {NoStop}%
%
\bibitem [{\citenamefont {{M.~Kunitski et al.}}(2019)}]{ionisation}%
  \BibitemOpen
  \bibfield  {author} {\bibinfo {author} {\bibnamefont {{M.~Kunitski et
  al.}}},\ }\href {\doibase 10.1038/s41467-018-07882-8} {\bibfield  {journal}
  {\bibinfo  {journal} {Nature Comms.}\ }\textbf {\bibinfo {volume} {10}}
  (\bibinfo {year} {2019}),\ 10.1038/s41467-018-07882-8}\BibitemShut {NoStop}%
%
%
%
\bibitem [{\citenamefont {Heinzl}\ \emph {et~al.}(2010)\citenamefont {Heinzl},
  \citenamefont {Ilderton},\ and\ \citenamefont {Marklund}}]{Heinzl:2010vg}%
  \BibitemOpen
  \bibfield  {author} {\bibinfo {author} {\bibfnamefont {T.}~\bibnamefont
  {Heinzl}}, \bibinfo {author} {\bibfnamefont {A.}~\bibnamefont {Ilderton}}, \
  and\ \bibinfo {author} {\bibfnamefont {M.}~\bibnamefont {Marklund}},\ }\href
  {\doibase 10.1016/j.physletb.2010.07.044} {\bibfield  {journal} {\bibinfo
  {journal} {Phys. Lett.}\ }\textbf {\bibinfo {volume} {B692}},\ \bibinfo
  {pages} {250} (\bibinfo {year} {2010})},\ \Eprint
  {http://arxiv.org/abs/1002.4018} {arXiv:1002.4018 [hep-ph]} \BibitemShut
  {NoStop}%
%
\bibitem{RES}
A.A.~Gonoskov, A.V.~Korzhimanov, A.V.~Kim, M.~Marklund and A.M.~Sergeev,
\href{https://link.aps.org/doi/10.1103/PhysRevE.84.046403}{Phys. Rev. E {\bf 84} (2011) 046403}
%
\bibitem{XUV}
T.G.~Blackburn, A.A.~Gonoskov and M.~Marklund,
\href{https://link.aps.org/doi/10.1103/PhysRevA.98.023421}{Phys. Rev. A {\bf 98} (2018) 023421}
%
\bibitem [{\citenamefont {Burke}\ \emph {et~al.}(1997)\citenamefont {Burke},
  \citenamefont {Field}, \citenamefont {Horton-Smith}, \citenamefont {Spencer},
  \citenamefont {Walz}, \citenamefont {Berridge}, \citenamefont {Bugg},
  \citenamefont {Shmakov}, \citenamefont {Weidemann}, \citenamefont {Bula},
  \citenamefont {McDonald}, \citenamefont {Prebys}, \citenamefont {Bamber},
  \citenamefont {Boege}, \citenamefont {Koffas}, \citenamefont {Kotseroglou},
  \citenamefont {Melissinos}, \citenamefont {Meyerhofer}, \citenamefont
  {Reis},\ and\ \citenamefont {Ragg}}]{SLAC-E144}%
  \BibitemOpen
  \bibfield  {author} {\bibinfo {author} {\bibfnamefont {D.~L.}\ \bibnamefont
  {Burke}}, \bibinfo {author} {\bibfnamefont {R.~C.}\ \bibnamefont {Field}},
  \bibinfo {author} {\bibfnamefont {G.}~\bibnamefont {Horton-Smith}}, \bibinfo
  {author} {\bibfnamefont {J.~E.}\ \bibnamefont {Spencer}}, \bibinfo {author}
  {\bibfnamefont {D.}~\bibnamefont {Walz}}, \bibinfo {author} {\bibfnamefont
  {S.~C.}\ \bibnamefont {Berridge}}, \bibinfo {author} {\bibfnamefont {W.~M.}\
  \bibnamefont {Bugg}}, \bibinfo {author} {\bibfnamefont {K.}~\bibnamefont
  {Shmakov}}, \bibinfo {author} {\bibfnamefont {A.~W.}\ \bibnamefont
  {Weidemann}}, \bibinfo {author} {\bibfnamefont {C.}~\bibnamefont {Bula}},
  \bibinfo {author} {\bibfnamefont {K.~T.}\ \bibnamefont {McDonald}}, \bibinfo
  {author} {\bibfnamefont {E.~J.}\ \bibnamefont {Prebys}}, \bibinfo {author}
  {\bibfnamefont {C.}~\bibnamefont {Bamber}}, \bibinfo {author} {\bibfnamefont
  {S.~J.}\ \bibnamefont {Boege}}, \bibinfo {author} {\bibfnamefont
  {T.}~\bibnamefont {Koffas}}, \bibinfo {author} {\bibfnamefont
  {T.}~\bibnamefont {Kotseroglou}}, \bibinfo {author} {\bibfnamefont {A.~C.}\
  \bibnamefont {Melissinos}}, \bibinfo {author} {\bibfnamefont {D.~D.}\
  \bibnamefont {Meyerhofer}}, \bibinfo {author} {\bibfnamefont {D.~A.}\
  \bibnamefont {Reis}}, \ and\ \bibinfo {author} {\bibfnamefont
  {W.}~\bibnamefont {Ragg}},\ }\href {\doibase 10.1103/PhysRevLett.79.1626}
  {\bibfield  {journal} {\bibinfo  {journal} {Phys. Rev. Lett.}\ }\textbf
  {\bibinfo {volume} {79}},\ \bibinfo {pages} {1626} (\bibinfo {year}
  {1997})}\BibitemShut {NoStop}%
\bibitem [{\citenamefont {Sch{\"u}tzhold}\ \emph {et~al.}(2008)\citenamefont
  {Sch{\"u}tzhold}, \citenamefont {Gies},\ and\ \citenamefont
  {Dunne}}]{Schutzhold:2008pz}%
  \BibitemOpen
  \bibfield  {author} {\bibinfo {author} {\bibfnamefont {R.}~\bibnamefont
  {Sch{\"u}tzhold}}, \bibinfo {author} {\bibfnamefont {H.}~\bibnamefont
  {Gies}}, \ and\ \bibinfo {author} {\bibfnamefont {G.}~\bibnamefont {Dunne}},\
  }\href {\doibase 10.1103/PhysRevLett.101.130404} {\bibfield  {journal}
  {\bibinfo  {journal} {Phys. Rev. Lett.}\ }\textbf {\bibinfo {volume} {101}},\
  \bibinfo {pages} {130404} (\bibinfo {year} {2008})},\ \Eprint
  {http://arxiv.org/abs/0807.0754} {arXiv:0807.0754 [hep-th]} \BibitemShut
  {NoStop}%
\bibitem [{\citenamefont {Dunne}\ \emph {et~al.}(2009)\citenamefont {Dunne},
  \citenamefont {Gies},\ and\ \citenamefont {Sch{\"u}tzhold}}]{Dunne:2009gi}%
  \BibitemOpen
  \bibfield  {author} {\bibinfo {author} {\bibfnamefont {G.~V.}\ \bibnamefont
  {Dunne}}, \bibinfo {author} {\bibfnamefont {H.}~\bibnamefont {Gies}}, \ and\
  \bibinfo {author} {\bibfnamefont {R.}~\bibnamefont {Sch{\"u}tzhold}},\ }\href
  {\doibase 10.1103/PhysRevD.80.111301} {\bibfield  {journal} {\bibinfo
  {journal} {Phys. Rev.}\ }\textbf {\bibinfo {volume} {D80}},\ \bibinfo {pages}
  {111301} (\bibinfo {year} {2009})},\ \Eprint {http://arxiv.org/abs/0908.0948}
  {arXiv:0908.0948 [hep-ph]} \BibitemShut {NoStop}%
\bibitem [{\citenamefont {Di~Piazza}\ \emph {et~al.}(2009)\citenamefont
  {Di~Piazza}, \citenamefont {Lotstedt}, \citenamefont {Milstein},\ and\
  \citenamefont {Keitel}}]{DiPiazza:2009py}%
  \BibitemOpen
  \bibfield  {author} {\bibinfo {author} {\bibfnamefont {A.}~\bibnamefont
  {Di~Piazza}}, \bibinfo {author} {\bibfnamefont {E.}~\bibnamefont {Lotstedt}},
  \bibinfo {author} {\bibfnamefont {A.~I.}\ \bibnamefont {Milstein}}, \ and\
  \bibinfo {author} {\bibfnamefont {C.~H.}\ \bibnamefont {Keitel}},\ }\href
  {\doibase 10.1103/PhysRevLett.103.170403} {\bibfield  {journal} {\bibinfo
  {journal} {Phys. Rev. Lett.}\ }\textbf {\bibinfo {volume} {103}},\ \bibinfo
  {pages} {170403} (\bibinfo {year} {2009})},\ \Eprint
  {http://arxiv.org/abs/0906.0726} {arXiv:0906.0726 [hep-ph]} \BibitemShut
  {NoStop}%
\bibitem [{\citenamefont {Bulanov}\ \emph {et~al.}(2010)\citenamefont
  {Bulanov}, \citenamefont {Mur}, \citenamefont {Narozhny}, \citenamefont
  {Nees},\ and\ \citenamefont {Popov}}]{Bulanov:2010ei}%
  \BibitemOpen
  \bibfield  {author} {\bibinfo {author} {\bibfnamefont {S.~S.}\ \bibnamefont
  {Bulanov}}, \bibinfo {author} {\bibfnamefont {V.~D.}\ \bibnamefont {Mur}},
  \bibinfo {author} {\bibfnamefont {N.~B.}\ \bibnamefont {Narozhny}}, \bibinfo
  {author} {\bibfnamefont {J.}~\bibnamefont {Nees}}, \ and\ \bibinfo {author}
  {\bibfnamefont {V.~S.}\ \bibnamefont {Popov}},\ }\href {\doibase
  10.1103/PhysRevLett.104.220404} {\bibfield  {journal} {\bibinfo  {journal}
  {Phys. Rev. Lett.}\ }\textbf {\bibinfo {volume} {104}},\ \bibinfo {pages}
  {220404} (\bibinfo {year} {2010})},\ \Eprint {http://arxiv.org/abs/1003.2623}
  {arXiv:1003.2623 [hep-ph]} \BibitemShut {NoStop}%
\bibitem [{\citenamefont {Kohlfurst}\ \emph {et~al.}(2013)\citenamefont
  {Kohlfurst}, \citenamefont {Mitter}, \citenamefont {von Winckel},
  \citenamefont {Hebenstreit},\ and\ \citenamefont
  {Alkofer}}]{Kohlfurst:2012rb}%
  \BibitemOpen
  \bibfield  {author} {\bibinfo {author} {\bibfnamefont {C.}~\bibnamefont
  {Kohlfurst}}, \bibinfo {author} {\bibfnamefont {M.}~\bibnamefont {Mitter}},
  \bibinfo {author} {\bibfnamefont {G.}~\bibnamefont {von Winckel}}, \bibinfo
  {author} {\bibfnamefont {F.}~\bibnamefont {Hebenstreit}}, \ and\ \bibinfo
  {author} {\bibfnamefont {R.}~\bibnamefont {Alkofer}},\ }\href {\doibase
  10.1103/PhysRevD.88.045028} {\bibfield  {journal} {\bibinfo  {journal} {Phys.
  Rev.}\ }\textbf {\bibinfo {volume} {D88}},\ \bibinfo {pages} {045028}
  (\bibinfo {year} {2013})},\ \Eprint {http://arxiv.org/abs/1212.1385}
  {arXiv:1212.1385 [hep-ph]} \BibitemShut {NoStop}%
\bibitem [{\citenamefont {Gonoskov}\ \emph {et~al.}(2013)\citenamefont
  {Gonoskov}, \citenamefont {Gonoskov}, \citenamefont {Harvey}, \citenamefont
  {Ilderton}, \citenamefont {Kim}, \citenamefont {Marklund}, \citenamefont
  {Mourou},\ and\ \citenamefont {Sergeev}}]{Gonoskov:2013ada}%
  \BibitemOpen
  \bibfield  {author} {\bibinfo {author} {\bibfnamefont {A.}~\bibnamefont
  {Gonoskov}}, \bibinfo {author} {\bibfnamefont {I.}~\bibnamefont {Gonoskov}},
  \bibinfo {author} {\bibfnamefont {C.}~\bibnamefont {Harvey}}, \bibinfo
  {author} {\bibfnamefont {A.}~\bibnamefont {Ilderton}}, \bibinfo {author}
  {\bibfnamefont {A.}~\bibnamefont {Kim}}, \bibinfo {author} {\bibfnamefont
  {M.}~\bibnamefont {Marklund}}, \bibinfo {author} {\bibfnamefont
  {G.}~\bibnamefont {Mourou}}, \ and\ \bibinfo {author} {\bibfnamefont {A.~M.}\
  \bibnamefont {Sergeev}},\ }\href {\doibase 10.1103/PhysRevLett.111.060404}
  {\bibfield  {journal} {\bibinfo  {journal} {Phys. Rev. Lett.}\ }\textbf
  {\bibinfo {volume} {111}},\ \bibinfo {pages} {060404} (\bibinfo {year}
  {2013})},\ \Eprint {http://arxiv.org/abs/1302.4653} {arXiv:1302.4653
  [hep-ph]} \BibitemShut {NoStop}%
\bibitem [{\citenamefont {Hebenstreit}\ and\ \citenamefont
  {Fillion-Gourdeau}(2014)}]{Hebenstreit:2014lra}%
  \BibitemOpen
  \bibfield  {author} {\bibinfo {author} {\bibfnamefont {F.}~\bibnamefont
  {Hebenstreit}}\ and\ \bibinfo {author} {\bibfnamefont {F.}~\bibnamefont
  {Fillion-Gourdeau}},\ }\href {\doibase 10.1016/j.physletb.2014.10.056}
  {\bibfield  {journal} {\bibinfo  {journal} {Phys. Lett.}\ }\textbf {\bibinfo
  {volume} {B739}},\ \bibinfo {pages} {189} (\bibinfo {year} {2014})},\ \Eprint
  {http://arxiv.org/abs/1409.7943} {arXiv:1409.7943 [hep-ph]} \BibitemShut
  {NoStop}%
\bibitem [{\citenamefont {Abramowicz}\ \emph {et~al.}(2019)\citenamefont
  {Abramowicz} \emph {et~al.}}]{Abramowicz:2019gvx}%
  \BibitemOpen
  \bibfield  {author} {\bibinfo {author} {\bibfnamefont {H.}~\bibnamefont
  {Abramowicz}} \emph {et~al.},\ }\href@noop {} {\  (\bibinfo {year} {2019})},\
  \Eprint {http://arxiv.org/abs/1909.00860} {arXiv:1909.00860
  [physics.ins-det]} \BibitemShut {NoStop}%
\bibitem [{E32()}]{E320}%
  \BibitemOpen
  \href@noop {} {\ }\bibinfo {note} {{The E-320 experiment at
  SLAC}}\BibitemShut {NoStop}%
%
%
\bibitem{Jag}
  A.~Ilderton,
  \href{\doibase 10.1103/PhysRevD.100.125018}
  {Phys.\ Rev.\ D {\bf 100} (2019) no.12,  125018},
  \Eprint{http://arxiv.org/abs/1909.02484}{arXiv:1909.02484 [hep-ph]}
%
\bibitem [{\citenamefont {Dinu}\ \emph {et~al.}(2012)\citenamefont {Dinu},
  \citenamefont {Heinzl},\ and\ \citenamefont {Ilderton}}]{Dinu:2012tj}%
  \BibitemOpen
  \bibfield  {author} {\bibinfo {author} {\bibfnamefont {V.}~\bibnamefont
  {Dinu}}, \bibinfo {author} {\bibfnamefont {T.}~\bibnamefont {Heinzl}}, \ and\
  \bibinfo {author} {\bibfnamefont {A.}~\bibnamefont {Ilderton}},\ }\href
  {\doibase 10.1103/PhysRevD.86.085037} {\bibfield  {journal} {\bibinfo
  {journal} {Phys. Rev.}\ }\textbf {\bibinfo {volume} {D86}},\ \bibinfo {pages}
  {085037} (\bibinfo {year} {2012})},\ \Eprint {http://arxiv.org/abs/1206.3957}
  {arXiv:1206.3957 [hep-ph]} \BibitemShut {NoStop}%
\bibitem [{\citenamefont {Ritus}(1985)}]{RitusReview}%
  \BibitemOpen
  \bibfield  {author} {\bibinfo {author} {\bibfnamefont {V.~I.}\ \bibnamefont
  {Ritus}},\ }\href@noop {} {\bibfield  {journal} {\bibinfo  {journal} {J.
  Russ. Laser Res.}\ }\textbf {\bibinfo {volume} {6}},\ \bibinfo {pages} {497}
  (\bibinfo {year} {1985})}\BibitemShut {NoStop}%
\bibitem [{\citenamefont {Di~Piazza}\ \emph {et~al.}(2012)\citenamefont
  {Di~Piazza}, \citenamefont {Muller}, \citenamefont {Hatsagortsyan},\ and\
  \citenamefont {Keitel}}]{DiPiazza:2011tq}%
  \BibitemOpen
  \bibfield  {author} {\bibinfo {author} {\bibfnamefont {A.}~\bibnamefont
  {Di~Piazza}}, \bibinfo {author} {\bibfnamefont {C.}~\bibnamefont {Muller}},
  \bibinfo {author} {\bibfnamefont {K.~Z.}\ \bibnamefont {Hatsagortsyan}}, \
  and\ \bibinfo {author} {\bibfnamefont {C.~H.}\ \bibnamefont {Keitel}},\
  }\href {\doibase 10.1103/RevModPhys.84.1177} {\bibfield  {journal} {\bibinfo
  {journal} {Rev. Mod. Phys.}\ }\textbf {\bibinfo {volume} {84}},\ \bibinfo
  {pages} {1177} (\bibinfo {year} {2012})},\ \Eprint
  {http://arxiv.org/abs/1111.3886} {arXiv:1111.3886 [hep-ph]} \BibitemShut
  {NoStop}%
\bibitem{volkov35}
	D.~M.~Volkov, Z.~Phys.~{\bf 94} (1935) 250.
  %
\bibitem [{\citenamefont {Kibble}(1965)}]{Kibble:1965zza}%
  \BibitemOpen
  \bibfield  {author} {\bibinfo {author} {\bibfnamefont {T.~W.~B.}\
  \bibnamefont {Kibble}},\ }\href {\doibase 10.1103/PhysRev.138.B740}
  {\bibfield  {journal} {\bibinfo  {journal} {Phys. Rev.}\ }\textbf {\bibinfo
  {volume} {138}},\ \bibinfo {pages} {B740} (\bibinfo {year}
  {1965})}\BibitemShut {NoStop}%
\bibitem [{\citenamefont {Peskin}\ and\ \citenamefont
  {Schroeder}(1995)}]{Peskin:1995ev}%
  \BibitemOpen
  \bibfield  {author} {\bibinfo {author} {\bibfnamefont {M.~E.}\ \bibnamefont
  {Peskin}}\ and\ \bibinfo {author} {\bibfnamefont {D.~V.}\ \bibnamefont
  {Schroeder}},\ }\href {http://www.slac.stanford.edu/~mpeskin/QFT.html} {\emph
  {\bibinfo {title} {{An Introduction to quantum field theory}}}}\ (\bibinfo
  {publisher} {Addison-Wesley},\ \bibinfo {address} {Reading, USA},\ \bibinfo
  {year} {1995})\BibitemShut {NoStop}%
\bibitem [{\citenamefont {Seipt}(2017)}]{Seipt:2017ckc}%
  \BibitemOpen
  \bibfield  {author} {\bibinfo {author} {\bibfnamefont {D.}~\bibnamefont
  {Seipt}},\ }in\ \href {\doibase 10.3204/DESY-PROC-2016-04/Seipt} {\emph
  {\bibinfo {booktitle} {{Proceedings, HQ 2016: Dubna, Russia, July 18-30,
  2016}}}}\ (\bibinfo {year} {2017})\ pp.\ \bibinfo {pages} {24--43},\ \Eprint
  {http://arxiv.org/abs/1701.03692} {arXiv:1701.03692 [physics.plasm-ph]}
  \BibitemShut {NoStop}%
\bibitem [{\citenamefont {Dirac}(1955)}]{Dirac:1955uv}%
  \BibitemOpen
  \bibfield  {author} {\bibinfo {author} {\bibfnamefont {P.~A.~M.}\
  \bibnamefont {Dirac}},\ }\href {\doibase 10.1139/p55-081} {\bibfield
  {journal} {\bibinfo  {journal} {Can. J. Phys.}\ }\textbf {\bibinfo {volume}
  {33}},\ \bibinfo {pages} {650} (\bibinfo {year} {1955})}\BibitemShut
  {NoStop}%
\bibitem [{\citenamefont {Brodsky}\ \emph {et~al.}(1998)\citenamefont
  {Brodsky}, \citenamefont {Pauli},\ and\ \citenamefont
  {Pinsky}}]{Brodsky:1997de}%
  \BibitemOpen
  \bibfield  {author} {\bibinfo {author} {\bibfnamefont {S.~J.}\ \bibnamefont
  {Brodsky}}, \bibinfo {author} {\bibfnamefont {H.-C.}\ \bibnamefont {Pauli}},
  \ and\ \bibinfo {author} {\bibfnamefont {S.~S.}\ \bibnamefont {Pinsky}},\
  }\href {\doibase 10.1016/S0370-1573(97)00089-6} {\bibfield  {journal}
  {\bibinfo  {journal} {Phys. Rept.}\ }\textbf {\bibinfo {volume} {301}},\
  \bibinfo {pages} {299} (\bibinfo {year} {1998})},\ \Eprint
  {http://arxiv.org/abs/hep-ph/9705477} {arXiv:hep-ph/9705477 [hep-ph]}
  \BibitemShut {NoStop}%
\bibitem [{\citenamefont {Heinzl}(2001)}]{Heinzl:2000ht}%
  \BibitemOpen
  \bibfield  {author} {\bibinfo {author} {\bibfnamefont {T.}~\bibnamefont
  {Heinzl}},\ }\bibfield  {booktitle} {\emph {\bibinfo {booktitle} {{Methods of
  quantization}}},\ }\href {\doibase 10.1007/3-540-45114-5_2} {\bibfield
  {journal} {\bibinfo  {journal} {Lect. Notes Phys.}\ }\textbf {\bibinfo
  {volume} {572}},\ \bibinfo {pages} {55} (\bibinfo {year} {2001})},\ \Eprint
  {http://arxiv.org/abs/hep-th/0008096} {arXiv:hep-th/0008096 [hep-th]}
  \BibitemShut {NoStop}%
\bibitem [{not()}]{note1}%
  \BibitemOpen
  \href@noop {} {\ }\bibinfo {note} {{As a check, observe from (\ref{FFF}) that
  if we bring the two pulses together, $\Theta\to 0$, then we recover the
  single-pulse result with amplitude $\xi$; this is consistent, as in this
  limit the delta pulses combine into a single peak doing work
  $\xi$.}}\BibitemShut {Stop}%
\bibitem [{\citenamefont {Xia}\ \emph {et~al.}(2018)\citenamefont {Xia},
  \citenamefont {Tao}, \citenamefont {Cai}, \citenamefont {Fu},\ and\
  \citenamefont {Liu}}]{Glory}%
  \BibitemOpen
  \bibfield  {author} {\bibinfo {author} {\bibfnamefont {Q.~Z.}\ \bibnamefont
  {Xia}}, \bibinfo {author} {\bibfnamefont {J.~F.}\ \bibnamefont {Tao}},
  \bibinfo {author} {\bibfnamefont {J.}~\bibnamefont {Cai}}, \bibinfo {author}
  {\bibfnamefont {L.~B.}\ \bibnamefont {Fu}}, \ and\ \bibinfo {author}
  {\bibfnamefont {J.}~\bibnamefont {Liu}},\ }\href {\doibase
  10.1103/PhysRevLett.121.143201} {\bibfield  {journal} {\bibinfo  {journal}
  {Phys. Rev. Lett.}\ }\textbf {\bibinfo {volume} {121}},\ \bibinfo {pages}
  {143201} (\bibinfo {year} {2018})}\BibitemShut {NoStop}%
\bibitem{Dong:2017vse}
  S.~S.~Dong, M.~Chen, Q.~Su and R.~Grobe,
  \href{https://journals.aps.org/pra/abstract/10.1103/PhysRevA.96.032120}{Phys.\ Rev.\ A {\bf 96} (2017) 032120}

\bibitem [{\citenamefont {Huismans}\ \emph {et~al.}(2012)\citenamefont
  {Huismans}, \citenamefont {Gijsbertsen}, \citenamefont {Smolkowska},
  \citenamefont {Jungmann}, \citenamefont {Rouz\'ee}, \citenamefont {Logman},
  \citenamefont {L\'epine}, \citenamefont {Cauchy}, \citenamefont {Zamith},
  \citenamefont {Marchenko}, \citenamefont {Bakker}, \citenamefont {Berden},
  \citenamefont {Redlich}, \citenamefont {van~der Meer}, \citenamefont
  {Ivanov}, \citenamefont {Yan}, \citenamefont {Bauer}, \citenamefont
  {Smirnova},\ and\ \citenamefont {Vrakking}}]{Huismans2}%
  \BibitemOpen
  \bibfield  {author} {\bibinfo {author} {\bibfnamefont {Y.}~\bibnamefont
  {Huismans}}, \bibinfo {author} {\bibfnamefont {A.}~\bibnamefont
  {Gijsbertsen}}, \bibinfo {author} {\bibfnamefont {A.~S.}\ \bibnamefont
  {Smolkowska}}, \bibinfo {author} {\bibfnamefont {J.~H.}\ \bibnamefont
  {Jungmann}}, \bibinfo {author} {\bibfnamefont {A.}~\bibnamefont {Rouz\'ee}},
  \bibinfo {author} {\bibfnamefont {P.~S. W.~M.}\ \bibnamefont {Logman}},
  \bibinfo {author} {\bibfnamefont {F.}~\bibnamefont {L\'epine}}, \bibinfo
  {author} {\bibfnamefont {C.}~\bibnamefont {Cauchy}}, \bibinfo {author}
  {\bibfnamefont {S.}~\bibnamefont {Zamith}}, \bibinfo {author} {\bibfnamefont
  {T.}~\bibnamefont {Marchenko}}, \bibinfo {author} {\bibfnamefont {J.~M.}\
  \bibnamefont {Bakker}}, \bibinfo {author} {\bibfnamefont {G.}~\bibnamefont
  {Berden}}, \bibinfo {author} {\bibfnamefont {B.}~\bibnamefont {Redlich}},
  \bibinfo {author} {\bibfnamefont {A.~F.~G.}\ \bibnamefont {van~der Meer}},
  \bibinfo {author} {\bibfnamefont {M.~Y.}\ \bibnamefont {Ivanov}}, \bibinfo
  {author} {\bibfnamefont {T.-M.}\ \bibnamefont {Yan}}, \bibinfo {author}
  {\bibfnamefont {D.}~\bibnamefont {Bauer}}, \bibinfo {author} {\bibfnamefont
  {O.}~\bibnamefont {Smirnova}}, \ and\ \bibinfo {author} {\bibfnamefont
  {M.~J.~J.}\ \bibnamefont {Vrakking}},\ }\href {\doibase
  10.1103/PhysRevLett.109.013002} {\bibfield  {journal} {\bibinfo  {journal}
  {Phys. Rev. Lett.}\ }\textbf {\bibinfo {volume} {109}},\ \bibinfo {pages}
  {013002} (\bibinfo {year} {2012})}\BibitemShut {NoStop}%
\bibitem [{\citenamefont {Huismans}\ \emph {et~al.}(2011)\citenamefont
  {Huismans}, \citenamefont {Rouzee}, \citenamefont {Gijsbertsen},
  \citenamefont {Jungmann}, \citenamefont {Smolkowska}, \citenamefont {Logman},
  \citenamefont {Lepine}, \citenamefont {Cauchy}, \citenamefont {Zamith},
  \citenamefont {Marchenko}, \citenamefont {Bakker}, \citenamefont {Berden},
  \citenamefont {Redlich}, \citenamefont {van~der Meer}, \citenamefont
  {Muller}, \citenamefont {Vermin}, \citenamefont {Schafer}, \citenamefont
  {Spanner}, \citenamefont {Ivanov}, \citenamefont {Smirnova}, \citenamefont
  {Bauer}, \citenamefont {Popruzhenko},\ and\ \citenamefont
  {Vrakking}}]{Huismans1}%
  \BibitemOpen
  \bibfield  {author} {\bibinfo {author} {\bibfnamefont {Y.}~\bibnamefont
  {Huismans}}, \bibinfo {author} {\bibfnamefont {A.}~\bibnamefont {Rouzee}},
  \bibinfo {author} {\bibfnamefont {A.}~\bibnamefont {Gijsbertsen}}, \bibinfo
  {author} {\bibfnamefont {J.~H.}\ \bibnamefont {Jungmann}}, \bibinfo {author}
  {\bibfnamefont {A.~S.}\ \bibnamefont {Smolkowska}}, \bibinfo {author}
  {\bibfnamefont {P.}~\bibnamefont {Logman}}, \bibinfo {author} {\bibfnamefont
  {F.}~\bibnamefont {Lepine}}, \bibinfo {author} {\bibfnamefont
  {C.}~\bibnamefont {Cauchy}}, \bibinfo {author} {\bibfnamefont
  {S.}~\bibnamefont {Zamith}}, \bibinfo {author} {\bibfnamefont
  {T.}~\bibnamefont {Marchenko}}, \bibinfo {author} {\bibfnamefont {J.~M.}\
  \bibnamefont {Bakker}}, \bibinfo {author} {\bibfnamefont {G.}~\bibnamefont
  {Berden}}, \bibinfo {author} {\bibfnamefont {B.}~\bibnamefont {Redlich}},
  \bibinfo {author} {\bibfnamefont {A.~F.~G.}\ \bibnamefont {van~der Meer}},
  \bibinfo {author} {\bibfnamefont {H.~G.}\ \bibnamefont {Muller}}, \bibinfo
  {author} {\bibfnamefont {W.}~\bibnamefont {Vermin}}, \bibinfo {author}
  {\bibfnamefont {K.~J.}\ \bibnamefont {Schafer}}, \bibinfo {author}
  {\bibfnamefont {M.}~\bibnamefont {Spanner}}, \bibinfo {author} {\bibfnamefont
  {M.~Y.}\ \bibnamefont {Ivanov}}, \bibinfo {author} {\bibfnamefont
  {O.}~\bibnamefont {Smirnova}}, \bibinfo {author} {\bibfnamefont
  {D.}~\bibnamefont {Bauer}}, \bibinfo {author} {\bibfnamefont {S.~V.}\
  \bibnamefont {Popruzhenko}}, \ and\ \bibinfo {author} {\bibfnamefont
  {M.~J.~J.}\ \bibnamefont {Vrakking}},\ }\href {\doibase
  10.1126/science.1198450} {\bibfield  {journal} {\bibinfo  {journal}
  {Science}\ }\textbf {\bibinfo {volume} {331}},\ \bibinfo {pages} {61}
  (\bibinfo {year} {2011})}\BibitemShut {NoStop}%
\bibitem [{\citenamefont {Porat}\ \emph {et~al.}(2018)\citenamefont {Porat},
  \citenamefont {Alon}, \citenamefont {Rozen}, \citenamefont {Pedatzur},
  \citenamefont {Kruger}, \citenamefont {Azoury}, \citenamefont {Natan},
  \citenamefont {Orenstein}, \citenamefont {Bruner}, \citenamefont {Vrakking},\
  and\ \citenamefont {Dudovich}}]{Porat}%
  \BibitemOpen
  \bibfield  {author} {\bibinfo {author} {\bibfnamefont {G.}~\bibnamefont
  {Porat}}, \bibinfo {author} {\bibfnamefont {G.}~\bibnamefont {Alon}},
  \bibinfo {author} {\bibfnamefont {S.}~\bibnamefont {Rozen}}, \bibinfo
  {author} {\bibfnamefont {O.}~\bibnamefont {Pedatzur}}, \bibinfo {author}
  {\bibfnamefont {M.}~\bibnamefont {Kruger}}, \bibinfo {author} {\bibfnamefont
  {D.}~\bibnamefont {Azoury}}, \bibinfo {author} {\bibfnamefont
  {A.}~\bibnamefont {Natan}}, \bibinfo {author} {\bibfnamefont
  {G.}~\bibnamefont {Orenstein}}, \bibinfo {author} {\bibfnamefont {B.~D.}\
  \bibnamefont {Bruner}}, \bibinfo {author} {\bibfnamefont {M.~J.~J.}\
  \bibnamefont {Vrakking}}, \ and\ \bibinfo {author} {\bibfnamefont
  {N.}~\bibnamefont {Dudovich}},\ }\href {\doibase 10.1038/s41467-018-05185-6}
  {\bibfield  {journal} {\bibinfo  {journal} {Nature Communications}\ }\textbf
  {\bibinfo {volume} {9}} (\bibinfo {year} {2018}),\
  10.1038/s41467-018-05185-6}\BibitemShut {NoStop}%
\bibitem{Harvey:2016wte}
  C.~Harvey, M.~Marklund and A.~R.~Holkundkar,
  \href{https://journals.aps.org/prab/abstract/10.1103/PhysRevAccelBeams.19.094701}{Phys.\ Rev.\ Accel.\ Beams {\bf 19} (2016) 094701}

\bibitem{Harvey:2016uiy}
  C.~Harvey, A.~Gonoskov, A.~Ilderton and M.~Marklund,
  Phys.\ Rev.\ Lett.\  {\bf 118} (2017) 105004


\bibitem{Bigourd}
 D.~Bigourd, B.~Chatel, W.P.~Schleich, B.~Girard,
 \href{https://journals.aps.org/prl/abstract/10.1103/PhysRevLett.100.030202}{Phys.\ Rev.\ Lett {\bf 100} (2008) 030202}

  
\end{thebibliography}
\end{document}